\documentstyle[12pt]{article}
\begin{document}
\baselineskip 18pt
\def\today{\ifcase\month\or
 January\or February\or March\or April\or May\or June\or
 July\or August\or September\or October\or November\or December\fi
 \space\number\day, \number\year}
\def\thebibliography#1{\section*{References\markboth
 {References}{References}}\list
% {[\arabic{enumi}]}{\settowidth\labelwidth{[#1]}
 \leftmargin\labelwidth
% \advance\leftmargin\labelsep
 \usecounter{enumi}}
 \def\newblock{\hskip .11em plus .33em minus .07em}
 \sloppy
 \sfcode`\.=1000\relax
\renewcommand{\thefootnote}
{\fnsymbol{footnote}}
%
%
%\  \
\vskip 0.5 true cm 
\begin{center}

{\large  
Baryon asymmetry and neutron EDM
}
\vskip 1.0 true cm

Mayumi Aoki$^1$, Noriyuki Oshimo$^2$, and Akio Sugamoto$^3$ \\ 
\smallskip
{\it $^1$Graduate School of Humanities and Sciences \\
Ochanomizu University, 
Bunkyo-ku, Tokyo 112-8610,  Japan \\} 
{\it $^2$Institute for Cosmic Ray Research \\
University of Tokyo, Tanashi, Tokyo 188-0002, Japan \\}
{\it $^3$Department of Physics \\
Ochanomizu University, Bunkyo-ku, Tokyo 112-8610, Japan \\}
\end{center}

\vskip 1.0 true cm

\def\gsim{{\mathop >\limits_\sim}}
\def\lsim{{\mathop <\limits_\sim}}
\def\r2{\sqrt 2}
\def\sw2{\sin^2\theta_W}
\def\v#1{v_#1}
\def\tb{\tan\beta}
\def\c2b{\cos 2\beta}
\def\sq{\tilde q}
\def\st{\tilde t}
\def\m#1{{\tilde m}_#1}
\def\mH{m_H}
\def\mg{{\tilde m}_g}
\def\M{\tilde M}
\def\mgr{m_{3/2}}
\def\dW{\delta_W}
\def\vW2{v_W^2}
\def\muB{\mu_B}

     A non-vanishing value of the neutron EDM (electric dipole moment) 
is an unambiguous sign of $CP$ violation.   
The interaction Hamiltonian for the EDM and the MDM 
(magnetic dipole moment) is written by 
$H_{int}=-\mu_n\mbox{\boldmath$\sigma$}\cdot{\bf H} - d_{n} 
\mbox{\boldmath$\sigma$}\cdot{\bf E}$, 
where the transformation properties of the spin 
$\mbox{\boldmath$\sigma$}$, the magnetic field {\bf H}, 
and the electric field {\bf E} under $P$ (parity:  $\bf{x}\rightarrow-\bf{x}$) 
and $T$ (time reversal: $t\rightarrow -t)$) transformations are 
$(+, +, -)$ and $(-, -, +)$, respectively.   The different properties between  
{\bf H} and {\bf E} under P and T come from 
the Maxwell equation ${\rm rot} \bf {E}= - \dot{\bf{H}}$. 
We can see that the MDM operator does not break $P$ and $T$, 
but the EDM operator breaks both.  

     Baryon asymmetry of our universe implies $CP$ violation. 
The number of baryons in the universe dominates over that of 
anti-baryons.  
This asymmetry is measured by the ratio of the baryon number 
density $n_{B}$ to the photon number density $n_{\gamma}$, 
which can be estimated as follows:  
We have galaxies separated by a mean distance of 
6 Mpc $= 1.8 \times 10^{25}$cm, whose masses are roughly given by 
$3\times10^{11}$ times the solar mass $2\times 10^{33}$g. 
This gives $n_{B}\sim 10^{-8}/$cm$^{3}$. 
We have cosmic background radiation of 2.7 K, from which 
the Stephan-Boltzmann formula gives 
$n_{\gamma}\sim400/$cm$^{3}$.  
Therefore, a very very rough estimate is given by 
$n_{B}/n_{\gamma}\sim 10^{-10}$.  
To generate baryon asymmetry from a baryon symmetric state at
a high temperature, we need different reaction rates 
between particles and anti-particles, 
which is realized by $CP$ violation.   

     Always we feel frustration at the smallness 
of $CP$ violation in the Standard Model (SM).  The neutron EDM
vanishes exactly at the one- and two-loop levels. 
Its value is predicted as $|d_{n}|_{SM}<10^{-31}e\cdot$cm, 
which is much smaller than the present experimental upper bound of 
$1.2\times10^{-25}e\cdot$cm.  
Possibly generated baryon asymmetry is at most  
$n_{B}/n_{\gamma}\sim 10^{-25}$, which is
much smaller than the observed value.  
Here, the asymmetry suffers from severe suppression
by mass differences of quarks 
$\prod_{1 \le i \ne j\le 3} (m_{u_{i}}^2 - m_{u_{j}}^2) 
(m_{d_{i}}^2 - m_{d_{j}}^2)$/(100 GeV)$^{12} \sim 10^{-18}$, 
since no $CP$ violation occurs in the SM when two
quarks have the same mass (GIM cancellation).   
A new source of $CP$ violation is necessary to generate 
baryon asymmetry. 

     A hope of baryogenesis is in a supersymmetric (SUSY) extension of 
the SM, which is one of the most plausible candidates for physics 
beyond the SM.   In SUSY models, spin 0 particles corresponding to 
quarks exist, called squarks.   
For each flavor, there are two types of squarks, L and R,   
which are the superpartners of the left-handed and right-handed quarks, 
respectively.  
Spin 1/2 particles, called gauginos and Higgsinos, also 
appear as superpartners of gauge and Higgs bosons.  
The number of Higgs doublets is two. 
The mass matrices of these SUSY particles could be complex, 
and thus become new sources of $CP$ violation.

     At the electroweak scale, supersymmetry does not hold rigorously, 
since SUSY particles have not been observed yet, 
indicating their masses heavier than their partners.  
In order for supersymmetry to be broken softly without loosing good features 
for divergence, supersymmetry breaking terms should have dimensional  
coefficients.  
Possible such soft-breaking terms are gaugino
mass terms, squark and slepton mass-squared terms, 
scalar three-point couplings, and scalar two-point couplings.  
We assume that the three gaugino masses $\m3$, $\m2$, and $\m1$ 
for SU(3), SU(2), and U(1) gauge groups are unified at the energy scale of 
around $10^{16}$ GeV, according to the plausible unification of the coupling
constants. Since $\tilde{m}_i/\alpha_i$ is renormalization group invariant,
$\m3:\m2:\m1=\alpha_3:\alpha_2:(5/3)\alpha_1 \simeq 9:3:1$ 
at the electroweak scale.   
The squarks are also assumed to have the 
same mass at the unification scale.  Then at the electroweak scale, 
the squark masses of the first two generations are not much 
different from each other, typically denoted by $\M$.  

     Gauginos for SU(2)$\times$U(1) and Higgsinos are mixed within charged and neutral sectors, 
resulting in mass eigenstates, called charginos and neutralinos. 
The mass matrix for charginos is given by 
\begin{equation}
    M^- = \left(\matrix{\m2 & -g\v1^*/\r2 \cr
                -g\v2^*/\r2 & \mH}        \right),  
\label{chargino mass}
\end{equation}
where $\v1$ and $\v2$ stand for the vacuum expectation
values of the Higgs bosons, 
and $\mH$ for the Higgsino mass parameter.  
The down-type quarks and up-type quarks are respectively  
given masses from $\v1$ and $\v2$,   
while $\sqrt{ |\v1|^{2} +|\v2|^{2} }$ being 246 GeV determined from 
the $W$-boson mass.   
The diagonal (1,1) and (2,2) elements come from, respectively, 
the soft-breaking gaugino mass term 
and the mixing term of the Higgs superfields in superpotential. 
The off-diagonal (1,2) and (2,1) elements are due to the 
supersymmetrization of the charged current interactions for Higgs bosons, 
in which the $W$ and charged Higgs bosons are replaced by the charged 
gauginos and Higgsinos, and the neutral Higgs bosons by 
vacuum expectation values.  
The mass matrix for neutralinos is given by $4\times$4 matrix, 
which consists of $\m1$ and the same parameters as for the charginos.   

     The four parameters $\m2$, $\v1$, $\v2$, and $\mH$ in 
Eq. (\ref{chargino mass}) have complex values in general.   
Although the first three parameters can be made 
real and positive by rephasing fermion fields, the complex phase $\theta$ 
of $\mH$ remains physical, which is one origin of $CP$ violation 
in SUSY models.  

     The squarks L and R are mixed in each flavor through the Yukawa couplings  
in superpotential and the soft-breaking three-point couplings, 
where the neutral Higgs bosons are replaced by their vacuum
expectation values. The coefficient $A$ of the scalar three-point couplings  
is generally complex and cannot be made real, since we have no
further rephasing freedom for squarks after the phases of quarks and gauginos  
are fixed. The complex phase $\alpha$ of $A$ is
another source of $CP$ violation.   
Note that squark mixings among different generations are not 
necessary for squark mass-squared matrices to induce $CP$ violation.  
We have now two peculiar CP phases $\alpha$ and $\theta$ in SUSY 
models.

     The neutron EDM is induced by one-loop diagrams in 
which the charginos, neutralinos, or gluinos are exchanged 
together with squarks [1,2,3].  
The chirality flip of the quark by the EDM operator is embodied   
by the gaugino and Higgsino mixings or the L- and R-squark mixings, 
both of which violate $CP$.  
If the $CP$ violating phase $\theta$ is not so small, $\theta \gsim 0.1$, 
then the chargino-loop diagrams dominantly contribute to the neutron EDM, 
predicting its magnitude to be around the present experimental bound,  
$|d_{n}| \lsim 1\times 10^{-25}e\cdot$cm.  The squark masses are 
constrained as $\M \gsim 1$ TeV [1]. 
On the other hand, if $\theta$ is small but $\alpha$ is not 
suppressed, $\theta \ll 1$, $\alpha \sim 1$, 
then the gluino-loop diagrams dominate. 
The neutron EDM around the experimental bound is again predicted, 
although the constraint on the squark masses are relaxed as  
$\tilde{M} \gsim 100$ GeV for $\tilde{m_{2} } \gsim 500$ GeV [3]. 
The phase $\theta$ also induces the EDM of the $W$ boson 
at the one-loop level, leading to the neutron EDM at the 
two loop level.  If $\theta$ is not suppressed, 
the neutron EDM can be as large as of order $10^{-26}e\cdot$cm,  
irrespectively of the squark masses [4].  

     The same new sources of $CP$ violation in SUSY models can generate 
baryon asymmetry [2,3].  
When the universe cools down to 100 or 200 GeV, electroweak phase transition
occurs.  If it is first order, bubbles of the broken phase nucleate. 
The Higgs potential $V_{T}(v)$ have a bump ($V_{T}> 0$) 
separating unbroken ($V_{T}=0$) and broken ($V_{T}< 0$) vacua.  
To have a quasi-stable bubble, sufficient surface tension is
required at the bubble wall, where the potential takes a bump value.  The
Higgs vacuum expectation value is proportional to the masses of fermions, 
so that the change of $v$ from the unbroken vacuum through the wall to the
broken vacuum forms a kind of potential barrier, giving as well 
position-dependent $CP$ violation.  
Consequently, for instance, a reflection by the "potential barrier" 
of a right-handed 
gaugino to a left-handed Higgsino and its $CP$ conjugate reflection 
have different rates, producing $Y$ (hypercharge) flux into the
unbroken phase, since gauginos and Higgsinos have different values of $Y$.
Then, the net $Y$ density in the unbroken phase induces a chemical 
potential for the baryon number $B$ (pressure to increase $B$).  
Under this pressure the sphaleron transition works and generates $B$ 
in the unbroken phase through electroweak anomaly.  
 
     We have obtained the following results [2,3]:  
If the $CP$ phase $\theta$ is not suppressed, $\theta \gsim 0.1$, 
the charginos can mediate baryogenesis, giving $n_{B}/s = 10^{-10}-10^{-11}$.   
Here entropy density $s$ ("massless degree of freedom" per
unit volume) is used instead of $n_{\gamma}$, since entropy is preserved
under the adiabatic expansion of the universe.  
For $\theta \ll 1$, $\alpha \sim 1$, the top squarks can mediate baryogenesis, 
resulting $n_{B}/s = 10^{-10}-10^{-11}$.  
For these parameter values, 
the neutron EDM is predicted to have a large magnitude.   
Roughly speaking, $n_{B}/s =10^{-10}-10^{-11}$ 
corresponds to $|d_{n}| = 10^{-24}-10^{-26}e\cdot$cm.   
Therefore, the SUSY models explaining properly the baryon asymmetry 
of our universe predict $|d_{n}| =10^{-26}-10^{-25}e\cdot$cm.  

     The predicted values of the neutron EDM are around the present 
experimental upper bound.  These values can be examined by 
improvement of the experiment by only one order of magnitude.  
Such an improvement will not be so difficult for ultra cold neutron 
experiments [5].  We are anxiously awaiting our experimental colleagues 
to accomplish the task.

\vspace{1 cm}

\begin{center}
References   
\end{center}

\noindent
[1]  Y. Kizukuri and N. Oshimo, {\it Phys. Rev.} D45 (1992) 1806, 
                                         $\it{ibid.}$ D46 (1992) 3025.  

\noindent
[2]  M. Aoki, N. Oshimo, and A. Sugamoto, 
                {\it Prog. Theor. Phys.} 98 (1997) 1179. 

\noindent
[3]  M. Aoki, A. Sugamoto, and N. Oshimo,
                {\it Prog. Theor. Phys.} 98 (1997) 1325. 

\noindent
[4]  T. Kadoyoshi and N. Oshimo, {\it Phys. Rev.} D55 (1997) 1481. 

\noindent
[5]  H. Yoshiki et al., {\it Phys. Rev. Lett.} 68 (1992) 1323.  

\end{document}